\documentclass[journal,twocolumn]{IEEEtran}
\IEEEoverridecommandlockouts
\usepackage{cite}
\usepackage{amsmath,amssymb,amsfonts,amsthm}
\usepackage{algorithmic}
\usepackage{graphicx}
\usepackage{textcomp}
\usepackage{xcolor}
\def\BibTeX{{\rm B\kern-.05em{\sc i\kern-.025em b}\kern-.08em
    T\kern-.1667em\lower.7ex\hbox{E}\kern-.125emX}}
\usepackage[utf8]{inputenc}

\newcommand{\defeq}{\overset{\text{def}}{=}}

\newtheorem{thm}{Theorem}
\newtheoremstyle{italic}
  {\topsep}
  {\topsep}
  {}
  {}
  {\itshape}
  {}
  {.5em}
  {\thmname{#1}\thmnumber{ #2}\thmnote{ (#3)}}
\theoremstyle{italic}
\newtheorem{defn}{Definition}

\theoremstyle{italic}
\newtheorem{propn}{Proposition}

\theoremstyle{italic}
\newtheorem{lemma}{Lemma}
\newtheorem{cor}{Corollary}
\newtheorem{rem}{Remark}

\title{Feedback Increases the Capacity of Queues with Bounded Service Times }
\author{K. R. Sahasranand and Aslan Tchamkerten\thanks{KRS is with the School of Technology and Computer Science at Tata Institute of Fundamental Research, Mumbai, India; this work was carried out when he was with T\'el\'ecom Paris, France.  Email: sahasranand@iisc.ac.in.

AT is with the Department of Communications and Electronics, T\'el\'ecom Paris, $91120$ Palaiseau, France. Email: aslan.tchamkerten@telecom-paris.fr. 

A shorter version of this paper comprising the  main results and proof sketches in the discrete setting was presented at the 2023 Information Theory and Applications Workshop, and at the (2023) IEEE International Symposium on Information Theory~\cite{krsat23isit}.}}

\begin{document}

\maketitle

\begin{abstract}
In the ``Bits Through Queues" paper~\cite{anantharam1996bits}, it was hypothesized that full feedback always increases the capacity of first-in-first-out queues, except when the service time distribution is memoryless. More recently, a non-explicit sufficient condition under which feedback increases capacity was provided~\cite{aptel2020bits}, along with simple examples of service times meeting this condition. While this condition yields examples where feedback is beneficial, it does not offer explicit structural properties of such service times.

In this paper, we show that full feedback increases capacity whenever the service time has bounded support. This is achieved by investigating a generalized notion of feedback, with full feedback and weak feedback (introduced in~\cite{aptel2020bits}) as particular cases. 
\end{abstract}

\section{Introduction}
Anantharam and Verd\'u~\cite{anantharam1996bits} proposed a single-server queue as a channel model for conveying information via timing. Under this model, the message to be conveyed is encoded into arrival instants of packets to a single server queue. Each packet is ``served'' for a random duration before departing. The decoder observes the departure times of all the packets and finally outputs an estimate of the message. When full feedback is available, the transmitter is causally revealed the times at which packets exit the queue, and can use this information to select subsequent packet arrival times.

Under the First-In-First-Out (FIFO) service policy, given a service time $S$ with mean $\mathbb{E}[S]$ and a fixed queue output rate $\lambda < 1/\mathbb{E}[S]$, the full feedback capacity $C_{\mathrm{F}}(\lambda)$ of the channel is given by
\begin{equation}
C_{\mathrm{F}}(\lambda) = \lambda \big[ \sup_{\underset{\mathbb{E}[W] \le \frac{1}{\lambda}-\frac{1}{\mu}}{W \ge 0:}} H\left(W+S\right)-H(S)\big], \nonumber
\end{equation}
where $\mu=1/\mathbb{E}[S]$ denotes the rate of the server, $H(\cdot)$ denotes entropy, and the $W$'s are independent of $S$. Interestingly, Anantharam and Verd\'u showed that when $S$ is exponentially distributed the full feedback capacity $C_{\mathrm{F}}(\lambda)$ is equal to the capacity $C(\lambda)$ of the channel without feedback. 
Moreover, they hypothesized that full feedback always increases capacity, except in cases where the service time is memoryless—specifically, exponential in the continuous-time setting and geometric in the discrete-time setting~\cite{bedekar1998information}. The question of whether there exist service times for which feedback increases capacity remained unresolved until recently. This problem was addressed in~\cite{aptel2020bits}, where it was demonstrated that full feedback increases capacity when the service time satisfies the following condition:
\begin{align}
\sup_{\underset{W(X)\defeq (X-S_1)+}{X\ge0:\mathbb{E}[W(X)] \le \frac{1}{\lambda}-\frac{1}{\mu}}} H\left(W(X) + S_2\right) < \sup_{\underset{\mathbb{E}[W] \le \frac{1}{\lambda}-\frac{1}{\mu}}{W \ge 0:}} H\left(W+S\right)
\label{eq:weakfb}
\end{align}
for any $\lambda < \mu$. Here, $S_1$, $S_2$, are independent copies of $S$, the $X$'s are independent of $(S_1, S_2)$, and the $W$'s are independent of $S$.  Inequality \eqref{eq:weakfb}
 was derived by considering a weaker form of feedback, where the transmitter is causally informed of the times at which packets enter service, rather than the times at which packets depart the server. If the capacity under weak feedback is strictly smaller than the capacity under full feedback, then full feedback increases capacity. The left-side of~\eqref{eq:weakfb} corresponds to an upper bound on the maximum output entropy under weak feedback, whereas the right-side corresponds to the maximum output entropy under full feedback.

Inequality~\eqref{eq:weakfb} involves two convex optimization problems and can, in principle, be verified by examining the Karush-Kuhn-Tucker (KKT) conditions of each problem with respect to specific service times. Examples for which~\eqref{eq:weakfb} holds are the two equal mass point service time $\mathbb{P}(S=1)=\mathbb{P}(S=2)=1/2$ in the discrete-time setting, and $S\sim \text{Uniform}[0,1]$ in the continuous-time setting. However, even for such service times, the KKT verification turns out to be tedious. Beyond simple examples like the ones above, no explicit characterization of service time distributions for which feedback increases capacity was provided.

In this paper, we show that full feedback increases capacity for any service time with bounded support. For continuous-time queues, this is obtained through a direct analysis of inequality \eqref{eq:weakfb}. In particular, we show that when the service time distribution has bounded support, for the output distributions corresponding to the two sides of~\eqref{eq:weakfb} to match, a certain compact operator with infinite-dimensional domain must be invertible which it is not. Unfortunately, this argument does not carry over to the discrete case. Besides, unlike in the continuous case, direct analysis of~\eqref{eq:weakfb} in the discrete case for $S$ with essential supremum $k$ entails solving a $k$-th order non-homogeneous recursive relation for the left-side term, which is cumbersome. Instead, we show that sufficient condition \eqref{eq:weakfb} generalizes to 
\begin{align}
    \sup_{\underset{T \defeq (S_1-\tau)_+}{\underset{W(X)\defeq (X-T)_+}{X\geq 0:\mathbb{E}[W(X)] \le \frac{1}{\lambda}-\frac{1}{\mu}}}} H\left(W(X)+S_2\right)< \sup_{\underset{\mathbb{E}[W] \le \frac{1}{\lambda}-\frac{1}{\mu}}{W \ge 0:}} H(W+S)\label{geneineq}
    \end{align}
    for any $0\leq \tau\leq \infty$. The left-side upper bounds the maximum output entropy under ``generalized feedback.'' This term turns out to be nondecreasing in $\tau$, equal to the weak feedback maximum output entropy for $\tau=0$, and equal to the full feedback maximum output entropy for $\tau=\infty$. Interestingly, if the service time has finite support, then by choosing $$\tau = k-1,$$ the non-homogeneous recursive relation alluded to above becomes tractable since its order reduces from $k$ to one, and indeed it can be verified that \eqref{geneineq} holds.  

By contrast, if the service time has unbounded support, the previous arguments fail since it can be shown that inequality~\eqref{geneineq} becomes equality for all $\tau \in [0,\infty]$, in both discrete-time and continuous-time settings. In particular, this disproves a conjecture made in~\cite{aptel2020bits} regarding the tightness of the weak feedback capacity upper bound.

\subsection*{Related work}
The discrete-time variant of~\cite{anantharam1996bits} was considered in~\cite{thomas1997shannon} and was studied in detail in~\cite{bedekar1998information} with an extension to multiple packets being served at a time. The zero-error capacity of a class of timing channels with finite support service times was investigated in~\cite{kovavcevic2014zero,kovavcevic2017zero}; indeed, the methods used are quite different due to the combinatorial nature of zero-error information theory. Upper and lower bounds on the capacity of timing channels with bounded service times in the continuous-time setting were derived in~\cite{sellke2007capacity}. Finite block length achievable rates for queues with geometrically distributed service times were provided in~\cite{riedl2011finite}. The capacity of discrete-time queues with Poisson inputs and Poisson outputs was considered in~\cite{prabhakar2003entropy}. Bounds on the capacity of queues with random arrivals were derived in~\cite{ferrante2016timing}. Practical error-correcting codes over timing channels with memoryless service times were discussed in~\cite{li2011memoryless}. 

Other variants of the queue channel, namely multiserver and bufferless queues were studied in~\cite{sundaresan2006capacity} and~\cite{tavan2013bits}, respectively.
Beyond capacity characterization, timing channels have been investigated with a focus on robust decoding~\cite{sundaresan2000robust}, sequential decoding~\cite{sundaresan2000sequential}, and reliability function~\cite{arikan2002reliability, wagner2005zero}. Example applications of timing channels include estimation and stabilization of linear systems~\cite{khojasteh2018estimating,khojasteh2019stabilizing}, secure and covert communication~\cite{dunn2009secure,mukherjee2016covert,biswas2017survey,dvorkind2018rate,ghassami2018covert,wagner2019degradedness,soltani2020fundamental}, energy harvesting systems~\cite{tutuncuoglu2017binary}, molecular communication~\cite{kovavcevic2014zero,gohari2016information}, and more recently, real-time monitoring of stochastic processes~\cite{yu2023real}.

The rest of the paper is organized as follows. Section~\ref{s:model} introduces the model and provides the context for our results which are presented in Section~\ref{s:results}. Proofs of the main results are provided in Section~\ref{s:proofs}. We conclude with a summary and future directions in Section~\ref{s:conclusion}.

\subsubsection*{Notation} For $x \in \mathbb{R}$ we use $(x)_+$ to denote $ \max\{x,0\}$. The probability law of a random variable $S$ is denoted $P_S$. The cumulative distribution function (cdf) of a random variable $S$ is denoted $F_S$.  If $S$ is a discrete random variable, we use $p_S$ to denote its probability mass function (pmf). If $S$ is a continuous random variable, we use $f_S$ to denote its probability density function (pdf). The  expectation of $S$ is denoted $\mathbb{E}[S]$. The shorthand notation $supp(S)$ denotes the support of the pmf or the pdf of $S$, depending on whether $S$ is discrete or continuous.  We use $\text{ess}\sup (S)$ to denote the essential supremum of $S$. Finally, we use $H(\cdot)$ to denote the entropy function (both discrete and continuous, unless stated otherwise) and $h_b(p)$ to denote the binary entropy $-p\log p-(1-p)\log (1-p)$ for $p \in (0,1)$.
\section{System Model and Preliminaries}
\label{s:model}
We follow the FIFO single-server queuing model developed in~\cite{anantharam1996bits, aptel2020bits}. In this model, upon the departure of a packet, the server selects a packet from the queue on a first-come-first-served basis, which immediately enters service, and departs from the server after a random service time. Specifically, a packet departs the queue at instant 
\begin{align}
    d&=a+\Delta+S \nonumber\\
    &=b+S, \nonumber
     \end{align} 
where $a$ denotes the instant the packet arrives in the queue, $\Delta$ denotes the time spent by the packet waiting in the queue, where $$b\defeq a+\Delta$$ denotes the time when the packet enters service, and the nonnegative random variable $S\sim P_S$ denotes the random service time. This variable has mean $\mathbb{E}[S] = 1/\mu$, where $\mu>0$ denotes the rate of the queue (in packets per unit time). Packets are assumed to experience independent and identically distributed (i.i.d.) service times.  

\begin{rem}
    If $a, \Delta$, and $S$ are nonnegative integer-valued random variables the setting is said to be discrete, and if any of $a, \Delta$, or $S$ takes values over $\mathbb{R}^+$ (at least one of which is not an integer), the setting is said to be continuous. Throughout this paper results hold in both settings unless stated otherwise.
\end{rem}

\begin{defn}[Timing code]
    An $(n,M_n,T_n,\epsilon_n)$-code for a timing channel used without feedback consists of
    \begin{itemize}
        \item $M_n$ messages where each message $u$ is encoded into $n$ arrival instants $0 \le a_1 \le a_2 \cdots \le a_n$. The $n$th departure from the queue occurs on average, over equally likely messages, no later than $T_n$---the queue is assumed to be initially empty.
         \item A decoder that observes the departures $d_1,\ldots,d_n$ and outputs an estimate of the transmitted message $\widehat{u}(d^n)$ such that the probability of error averaged over equally likely messages  satisfies
         \begin{align}
                \mathbb{P}[\widehat{u} \neq u] \le \epsilon_n. \nonumber
         \end{align}
    \end{itemize}
     An $(n,M_n,T_n,\epsilon_n)$-code is a full feedback code if  $a_i$ is a function of $u$ and $d^{i-1}$. An $(n,M_n,T_n,\epsilon_n)$-code is a weak feedback code if $a_i$ is a function of $u$ and $b^{i-1}$.
\end{defn}
The rate of an $(n,M_n,T_n,\epsilon_n)$-code is defined as
\begin{align}
\frac{\log_2 M_n}{T_n} \nonumber
\end{align}
in bits per unit time.
\begin{defn}[Capacity]\label{def:cap}
The capacity $C$ without feedback is the supremum of all $R$ for which, for all $\gamma > 0$, there exists a sequence of $(n,M_n,T_n,\epsilon_n)$-codes without feedback with $T_n \to \infty$ and $\epsilon_n \to 0$ as $n \to \infty$ and such that 
\begin{align}
\frac{\log_2 M_n}{T_n} > R-\gamma. \nonumber
\end{align}
for all $n$ sufficiently large. 

The full feedback capacity $C_{\mathrm{F}}$ and the weak feedback capacity $C_{\mathrm{WF}}$ are defined analogously. 
\end{defn}

\begin{defn}[Capacity at fixed output/departure rate]\label{def:caplam}
Given $\lambda >  0$, the capacity $C(\lambda)$ without feedback at output rate $\lambda$ is the supremum of all $R$ for which, for all $\gamma > 0$, there exists a sequence of $(n,M_n,n/\lambda,\epsilon_n)$-codes without feedback such that  
\begin{align}
\lambda \frac{\log_2 M_n}{n} > R-\gamma, \nonumber
\end{align}
where $\epsilon_n \to 0$ as $n \to \infty$.
The capacities $C_{\mathrm{F}}(\lambda)$ and $C_{\mathrm{WF}}(\lambda)$ at output rate $\lambda$ under full feedback and weak feedback, respectively, are defined similarly.
\end{defn}

The weak feedback capacity can obviously not be lower than the capacity without feedback. What is perhaps less obvious is that the full feedback capacity is at least as large as the weak feedback capacity (hence the term ``weak''). This follows from the fact that $b_i$ is a function of $\{d_j\}_{j\leq i-1}$ and $\{a_j\}_{j\leq i}$. Hence we have:
\begin{lemma}
    [\hspace{-.03cm}\cite{aptel2020bits}]
\label{prop:clam}
For $0< \lambda < \mu$, we have
\begin{align}
C(\lambda) \le C_{\mathrm{WF}}(\lambda) \le C_{\mathrm{F}}(\lambda). \nonumber
\end{align}
\end{lemma}
From this lemma and the relationship between capacity and capacity at fixed output rate (see \cite[Theorem~$1$]{anantharam1996bits}) we get:
\begin{propn}
\label{prop:Clam}
   A queue with service rate $\mu>0$ satisfies
   \begin{align}
     C=\sup_{0< \lambda<\mu}C(\lambda) &\leq C_{\mathrm{WF}}=\sup_{0< \lambda<\mu}C_{\mathrm{WF}}(\lambda) \nonumber\\
     &\leq C_{\mathrm{F}} = \sup_{0< \lambda<\mu} C_{\mathrm{F}}(\lambda).  \nonumber
   \end{align}
\end{propn}
It is known (see~\cite{anantharam1996bits}) that when full feedback is available, encoding strategies may, without loss of optimality (in terms of error probability for the same communication rate), be restricted to those for which $$a_i \ge d_{i-1},$$ {\it{i.e.}},  the transmitter awaits the full feedback information before sending the next packet which immediately enters service ($\Delta=0$). This, in turn, reduces the communication over the queue channel to communication over the additive and memoryless channel 
\begin{align}
    W_i \to W_i + S_i,   \nonumber
\end{align}
where $$W_i \defeq a_i - d_{i-1}$$ denotes the waiting time of the queue---between the $(i-1)$st and the $i$th packets. In other words, this channel transforms the amount of time the queue waits for the $i$th packet into the total amount of time the queue spends on the $i$th packet, waiting and serving. The following result follows:
\begin{propn}[\hspace{-.03cm}\cite{anantharam1996bits}]\label{prop:FF}
    For $0< \lambda < \mu$, we have
    \begin{align}
    C_{\mathrm{F}}(\lambda) = \lambda \sup_{\underset{\mathbb{E}[W] \le \frac{1}{\lambda}-\frac{1}{\mu}}{W \ge 0:}} I\left(W ; W+S\right), \label{fexpr}
    \end{align}
    \label{prop:CF}
where the optimization is over all nonnegative random variables $W$ independent of $S$.
\end{propn}
In \cite{aptel2020bits} it was shown that under weak feedback encoding strategies may be restricted to those for which $$a_i \ge b_{i-1}.$$ Hence, similarly to the full feedback case, the transmitter awaits the weak feedback information before sending the next packet. Accordingly, to a given transmitter signal
\begin{align}
X_i \defeq a_i - b_{i-1} \ge 0, \nonumber
\end{align}
corresponds the waiting time of the queue  $$W_i(X_i)\defeq (X_i-S_{i-1})_+.$$ This, in turn, yields the additive (possibly with memory) channel
\begin{align}
W_i(X_i) \to W_i(X_i)+S_i \nonumber
\end{align}
whose maximum mutual information upper bounds the weak feedback capacity:
\begin{propn}[\hspace{-.03cm}\cite{aptel2020bits}]
    For $0< \lambda < \mu$, we have
    \begin{align}\label{upb}
    C_{\mathrm{WF}}(\lambda) \le \lambda \sup_{\underset{W(X)\defeq (X-S_1)_+}{X\geq 0:\mathbb{E}[W(X)] \le \frac{1}{\lambda}-\frac{1}{\mu}}} I\left(W(X) ; W(X) + S_2\right),
    \end{align}
where $S_1$ and $S_2$ are i.i.d. $P_S$ with mean $1/\mu$, and the optimization is over all nonnegative random variables $X$ independent of $(S_1,S_2)$. 
\label{prop:CWF_UB}
\end{propn}

\begin{rem}
  Proposition~\ref{prop:CWF_UB} gives only an upper bound on the weak feedback capacity. This is by contrast with the full feedback capacity expression of Proposition~\ref{prop:FF} which corresponds to the capacity of the additive channel $W\to W+S$. Indeed, under weak feedback the outputs $\{W_i(X_i)+S_i\}_{i=1}^n$ are generally dependent since $W_i(X_i)$ depends on $S_{i-1}$.  Proposition~\ref{prop:CWF_UB} is obtained by upper bounding the output entropy $H(\{W_i(X_i)+S_i\}_{i=1}^n)$ by the sum of the individual output entropies $\sum_{i=1}^n H(W_i(X_i)+S_i)$. 
   \end{rem} 

Observe that the right-sides of~\eqref{fexpr} and \eqref{upb} are continuous functions of $\lambda$ over the open interval $(0,\mu)$. This follows from the fact that the functions 
\begin{align}
q \mapsto \sup_{\underset{\mathbb{E}[W] \le q}{W \ge 0:}} H (W+S) \nonumber
\end{align}
and
\begin{align}
q \mapsto \sup_{\underset{W(X)\defeq (X-S_1)_+}{X\geq 0:\mathbb{E}[W(X)] \le q}} H( W(X) + S_2) \nonumber
\end{align}
are nonnegative and concave over $\mathbb{R}^+$ (see, {\it{e.g.}},~\cite[proof of Lemma $1$]{aptel2020bits}). Observe also that the right-sides of \eqref{fexpr} and \eqref{upb} tend to zero as $\lambda \uparrow \mu$ and as $\lambda \downarrow 0$.\footnote{For the latter this follows by noting that the mutual information terms on the right-sides of \eqref{fexpr} and \eqref{upb} are upper-bounded by $1+\ln(1/\lambda)$, the entropy of an exponential random variable with mean $1/\lambda$ in the continuous case, and by $\frac{\lambda+1}{\lambda} h_b\left(\frac{\lambda}{\lambda+1}\right)$, the entropy of a geometric random variable with mean $1/\lambda$ in the discrete case.} Hence, the supremum over $\lambda$ of the right-sides of \eqref{fexpr} and \eqref{upb} are achieved for some $\lambda \in (0,\mu)$, and Propositions~\ref{prop:clam} and \ref{prop:CWF_UB} imply:
\begin{cor}\label{corupt}
For a queue with service rate $\mu~>~0$, the weak feedback capacity is strictly less than the full feedback capacity if, for any $0<\lambda < \mu$, 
\begin{equation}
\sup_{\underset{W(X)\defeq (X-S_1)_+}{X\geq 0:\mathbb{E}[W(X)] \le \frac{1}{\lambda}-\frac{1}{\mu}}} H\left(W(X) + S_2\right) < \sup_{\underset{\mathbb{E}[W] \le \frac{1}{\lambda}-\frac{1}{\mu}}{W \ge 0:}} H\left(W+S\right).
\label{eq:strictineq}    
\end{equation}
 
\end{cor}

\section{Results}
\label{s:results}
 Generalized feedback, g-feedback in short, is said to be available if the transmitter has causal access to
\begin{align}
g_i \defeq (b_i,c_i), \nonumber
\end{align}
where
\begin{align}
c_i &\defeq b_i + \min\{S_i,\tau\}= \min\{d_i,b_i+\tau\} \nonumber
\end{align}
for some known fixed constant $\tau \geq 0$ and $1 \le i \le n$. 
The transmitter first learns $b_i$, the time at which the packet enters service, and later, at time $c_i$, it learns either the departure time of the packet (if $S_i < \tau$) or that the packet is still in service (if $S_i\geq \tau$). Under g-feedback, the capacity and the capacity at fixed output rate are denoted $C_{\mathrm{F}}^\tau$ and $C_{\mathrm{F}}^\tau(\lambda)$, respectively. 

Since
    \begin{align}
    c_i &= \min\{d_i,b_i+\tau\}, \nonumber
    \end{align}
and $b_i\leq d_i$, weak feedback and full feedback correspond to the cases $\tau=0$ and $\tau=\infty$, respectively. Indeed, for $\tau=0$ we have $c_i=b_i$, and for $\tau=\infty$ the transmitter has access to $g_i=(b_i,d_i)$ which is equivalent to having access to $d_i$ only since $b_i$ is a function of $\{d_j\}_{j\leq i-1}$ and $\{a_j\}_{j\leq i}$. Hence,
\begin{align}
C_{\mathrm{F}}^0= C_{\mathrm{WF}} \qquad \text{and} \qquad C_{\mathrm{F}}^\infty  = C_{\mathrm{F}},    \nonumber
\end{align} 
and similarly for the capacity at fixed output rate.

It turns out that the coding schemes available under g-feedback with parameter $\tau_2$ include those available under g-feedback with parameter ${\tau_1}\leq \tau_2$, hence we have the following monotone capacity property:
\begin{propn}
\label{thm:sandwich}
For any $\lambda < \mu$, we have that $C_{\mathrm{F}}^\tau(\lambda)$ is nondecreasing in~$\tau$ over $[0,\infty]$.
\end{propn}
As an immediate consequence we have:
\begin{cor}\label{lecor}
For any $0 < \lambda < \mu$ and $0\leq \tau\leq \infty$,
$$C(\lambda)\leq C_{\mathrm{WF}}(\lambda)\defeq C_\mathrm{F}^0(\lambda)\leq C_\mathrm{F}^\tau(\lambda)\leq C_\mathrm{F}^\infty(\lambda) \defeq C_\mathrm{F}(\lambda). $$
\end{cor}

As we saw in Section~\ref{s:model}, if either full feedback or weak feedback is available, it is optimal to send a packet after receiving the feedback information: $a_i\geq d_{i-1}$ under full feedback and $a_i\geq b_{i-1}$ under weak feedback. This generalizes to g-feedback:
\begin{propn}
\label{thm:wlog}
Under g-feedback, for any $0\leq \tau\leq \infty$, encoding strategies may, without loss of optimality, be restricted to those where 
\begin{align}
a_i \ge c_{i-1} \qquad ~1 \le i \le n. \nonumber
\end{align}
\end{propn}

Following Proposition~\ref{thm:wlog}, define under g-feedback the transmitter signal 
\begin{align}
X_i \defeq a_i-c_{i-1} \ge 0. \nonumber
\end{align}
This yields the additive channel
\begin{align}
    W_i(X_i) \rightarrow W_i(X_i) + S_i \qquad 1 \le i \le n, \nonumber
\end{align}
where 
\begin{align}
W_i(X_i) \defeq (X_i - (S_{i-1}-\tau)_+)_+   \nonumber
\end{align}
denotes the waiting time of the queue. The next result generalizes Proposition~\ref{prop:CWF_UB} and is obtained through similar arguments:
\begin{propn}
\label{thm:TFUB}
For any $0 < \lambda < \mu$ and $0 \le \tau \le \infty$, we have
\begin{align}
C_{\mathrm{F}}^\tau(\lambda) \le \lambda \sup_{\underset{T \defeq (S_1-\tau)_+}{\underset{W(X)\defeq (X-T)_+}{X\geq 0:\mathbb{E}[W(X)] \le \frac{1}{\lambda}-\frac{1}{\mu}}}} I\left(W(X) ; W(X)+S_2\right),  \nonumber
\end{align}
where $S_1$ and $S_2$ are i.i.d. $P_S$ with mean $1/\mu$, and the optimization is over all nonnegative random variables $X$ independent of $(S_1,S_2)$. 
\end{propn}

Using similar arguments as for Corollary~\ref{corupt},
Corollary~\ref{lecor} together with Propositions~\ref{prop:CF} and \ref{thm:TFUB} implies:
\begin{cor}\label{lecor2}
For a queue with service rate $\mu~>~0$, the g-feedback capacity with $\tau\in [0,\infty)$ is strictly less than the full feedback capacity if, for any $\lambda\in (0,\mu)$,
\begin{align}\sup_{\underset{T \defeq (S_1-\tau)_+}{\underset{W(X)\defeq (X-T)_+}{X\geq 0:\mathbb{E}[W(X)] \le \frac{1}{\lambda}-\frac{1}{\mu}}}} H\left(W(X)+S_2\right)< \sup_{\underset{\mathbb{E}[W] \le \frac{1}{\lambda}-\frac{1}{\mu}}{W \ge 0:}} H(W+S).\label{eq:strictineq2}
    \end{align}
 \end{cor}
Notice that Corollary~\ref{lecor2} with $\tau=0$ reduces to Corollary~\ref{corupt}.

In the discrete-time setting, it can be shown that for a bounded support service time $S$ inequality~\eqref{eq:strictineq2} holds by choosing  $$\tau=\text{ess}\sup(S) -1.$$ This implies:
\begin{thm}
For a discrete-time queue with bounded support service time $S$ we have 
\[
C \le C_\mathrm{F}^\tau<C_\mathrm{F}
\]
with $\tau=\text{ess}\sup(S) -1$. In particular, full feedback increases capacity for bounded support service times.

\label{thm:tau}
\end{thm}
From Proposition~\ref{thm:sandwich} and Theorem~\ref{thm:tau} we get:
\begin{cor}
\label{cor:WFB}
    For a discrete-time queue with bounded support service time we have $C_\mathrm{WF}<C_\mathrm{F}$.
\end{cor}

In the continuous-time setting, it can be shown that for a bounded support service time inequality~\eqref{eq:strictineq} holds (equivalently,~\eqref{eq:strictineq2} with $\tau=0$). This implies:
\begin{thm}\label{thm:ctsbdd}
 For a continuous-time queue, the full feedback capacity is strictly larger than the weak feedback capacity in the following two cases:
    \begin{enumerate}
    \item[i.] $supp(S) = [a,b]$ where $a > 0$ and $b < \infty$.
    \item[ii.] $supp(S) = [a,b]$ where $a \ge 0$ and $b < \infty$ and $f_S(\cdot)$ is continuous on $[a,b]$.
    \end{enumerate}    
In particular, in the above two cases, full feedback increases capacity.
\end{thm}

Unfortunately, for service times with unbounded supports, for both the discrete-time and the continuous-time settings, Corollary~\ref{lecor2} is inconclusive since inequality~\eqref{eq:strictineq2} is never satisfied:
\begin{thm}
\label{thm:inf}
For unbounded support service times we have for any $\tau\in [0,\infty]$ and any $\lambda \in (0,\mu)$,
\begin{align}
\sup_{\underset{T \defeq (S_1-\tau)_+}{\underset{W(X)\defeq (X-T)_+}{X\geq 0:\mathbb{E}[W(X)] \le \frac{1}{\lambda}-\frac{1}{\mu}}}} H\left(W(X)+S_2\right)
&=\sup_{\underset{\mathbb{E}[W] \le \frac{1}{\lambda}-\frac{1}{\mu}}{W \ge 0:}} H\left(W+S\right).
\label{eq:eqinf}
\end{align}
(As before, $S_1, S_2, S$ are i.i.d. $P_S$ with mean $1/\mu$, $X$ is a nonnegative random variable independent of $(S_1,S_2)$, and $W$ is independent of $S$. The theorem holds in both the discrete-time and the continuous-time settings.) 
\end{thm}
A few comments are in order. 
   Identity~\eqref{eq:eqinf} may come as a surprise since the optimization on its left-side imposes the extra condition that $W$ should be of the form $(X-T)_+$ with $T \defeq (S_1-\tau)_+$. 
We also note that Theorem~\ref{thm:inf}  disproves a conjecture made in \cite{aptel2020bits} that~\eqref{eq:eqinf} with $\tau=0$ holds only for geometric service times in the discrete-time setting (or only for exponential service times in the continuous-time setting). Theorem~\ref{thm:inf} also implies that for an unbounded service time either the weak feedback capacity upper bound given in Proposition~\ref{prop:CWF_UB} is not tight and $C_{\mathrm{WF}}<C_{\mathrm{F}}$, or that it is tight and $C_{\mathrm{WF}}=C_{\mathrm{F}}$. 

We end this section by discussing an alternative to g-feedback. The astute reader may wonder why g-feedback involves both $b_i$ and $c_i$, as opposed to only $c_i$. Indeed, for the particular case $\tau=\infty$, we have $c_i=d_i$, and g-feedback reduces to full feedback where only $c_i$ is fed back since $b_i$ can be retrieved from $\{a_j\}_{j\leq i}$ and $\{d_j\}_{j\leq i-1}$.  

 In the definition of g-feedback, suppose we replace $g_i$ by $c_i$. In this case, it can be shown that  Proposition~\ref{thm:wlog}, Proposition~\ref{thm:TFUB}, and Corollary~\ref{lecor2} still hold, but it is unclear if Proposition~\ref{thm:sandwich} holds. In particular, it is unclear  how $C_\mathrm{WF}(\lambda)$ and $C_\mathrm{F}^\tau(\lambda)$ with $0<\tau<\infty$ are related, and therefore if Corollary~\ref{cor:WFB} holds. 

\section{Proofs}
\label{s:proofs}

We first prove Proposition~\ref{thm:wlog}, followed by Proposition~\ref{thm:sandwich}, which relies on Proposition~\ref{thm:wlog}. We then prove Proposition~\ref{thm:TFUB}, followed by Theorems~\ref{thm:tau}, \ref{thm:ctsbdd}, and \ref{thm:inf}.
\subsection{Proof of Proposition~\ref{thm:wlog}}
\label{p:wlog}

Fix a g-feedback encoding strategy $a_1,a_2,\ldots, a_n$ for message $u$. This induces a departure distribution $$\mathbb{P}(d^n|u)= \prod_i  \mathbb{P}(d_i|u,d^{i-1}). $$ We show that if we change each arrival $a_i$ with $$\widetilde{a}_i\defeq \max\{a_i,c_{i-1}\},$$ the conditional probabilities $\mathbb{P}(d_i|u,d^{i-1})$, $1\leq i\leq n$, remain the same, hence the input-output joint distribution $\mathbb{P}(u,d^n)$ remains the same, and so does the error probability (for any decoding rule). Towards this, first we show that 
\begin{align}
    \mathbb{P}(d_i|u,d^{i-1}) = \mathbb{P}(d_i|u,d^{i-1}, a^i, c^{i-1}).
    \label{eq:prop4first}
\end{align}
The symbol $a_1$ is uniquely determined by the message $u$ and we have 
\[
c_1 = \min\{d_1,a_1 + \tau\}
\]
as the queue is initially empty. For $i\ge 2$, the symbol $a_i$ depends on both $u$ and $c^{i-1}$ and from the definitions of $c_i$ and $b_i$, it can be seen that from $u$ and $d^{i-1}$, we can recursively compute $(a_j,c_j)$ for $j \le i-1$, and $a_i$ and hence~\eqref{eq:prop4first} follows. Thus,
\begin{align}
\mathbb{P}(d_i|u,d^{i-1}) &= \mathbb{P}(d_i|u,d^{i-1}, a^i, c^{i-1}) \nonumber\\
&= \mathbb{P}\left(S_i = d_i - \max\{a_i,d_{i-1}\}|u,d^{i-1}, a^i, c^{i-1}\right) \nonumber\\
&= \mathbb{P}\left(S_i = d_i - \max\{a_i,d_{i-1}\}\right) \nonumber\\
&= \mathbb{P}\left(S_i = d_i - \max\{\widetilde{a}_i,d_{i-1}\}\right). \nonumber
\end{align}
To justify the last step, observe that 
\begin{itemize}
    \item if $a_i \ge d_{i-1}$, then $a_i \ge c_{i-1}$ and hence $\widetilde{a}_i = a_i \ge d_{i-1}$; 
    \item if $a_i \le d_{i-1}$, then either $\widetilde{a}_i = a_i \le d_{i-1}$ or $\widetilde{a}_i = c_{i-1} \le d_{i-1}$, 
\end{itemize}
whereby
\[
\max\{a_i,d_{i-1}\} = \max\{\widetilde{a}_i,d_{i-1}\}.
\]
This completes the proof.

\subsection{Proof of Proposition~\ref{thm:sandwich}}
\label{p:sandwich}

We show that the coding schemes available under g-feedback strategy with parameter $\tau_2 < \infty$ include the coding schemes available under g-feedback strategy with parameter $\tau_1 \le \tau_2$. Denote the feedback corresponding to the $i$th packet under a g-feedback strategy with parameter $\tau_j$ by 
\begin{align}
g^{(j)}_i = (b_i,c^{(j)}_i), \nonumber
\end{align}
 where $c^{(j)}_i = \min\{d_i, b_i+\tau_j\}$. By Proposition~\ref{thm:wlog}, optimal policies under g-feedback with parameter $\tau_1$ satisfy
\begin{align}
a_i \ge c^{(1)}_{i-1}. \nonumber
\end{align}
Consider a g-feedback strategy with parameter $\tau_2 \ge \tau_1$. Given $g^{(2)}_i = (b_i,c^{(2)}_i)$, we have
\begin{align}
\label{eq:recurcgen}
    c^{(1)}_i = \min\{b_i + \tau_1, c^{(2)}_i\}.
\end{align}
Therefore, given $\{g^{(2)}_1,\ldots,g^{(2)}_{i-2}\}$, the encoder can compute $b^{i-2}$, and $\{c^{(1)}_1,\ldots,c^{(1)}_{i-2}\}$ by~\eqref{eq:recurcgen}. Then, the encoder computes the value of $c^{(1)}_{i-1}$ at time instant $c^{(1)}_{i-1}$ as follows. The encoder, having received a feedback at $b_{i-1}$, sets 
\begin{align}
c^{(1)}_{i-1} = c^{(2)}_{i-1} \nonumber
\end{align}
if the feedback $c^{(2)}_{i-1}$ is received before $b_{i-1}+\tau_1$; otherwise sets 
\begin{align}
c^{(1)}_{i-1} = b_{i-1}+\tau_1. \nonumber
\end{align}
Thus, if $a_i$ is a function of $(g^{(1)}_1,\ldots,g^{(1)}_{i-1})$, then using g-feedback with parameter $\tau_2$, the encoder can compute $a_i$ no later than $c^{(1)}_{i-1}$. Therefore, a g-feedback strategy with with parameter $\tau_2$ can ``simulate'' a g-feedback strategy with parameter $\tau_1 \le \tau_2$. \hfill $\square$

\subsection{Proof of Proposition~\ref{thm:TFUB}}
 The proof follows the proof of \cite[Theorem~2]{aptel2020bits} and amounts to replacing $D_i=(X_i-S_{i-1})_++S_i$ with $D_i=(X_i-T_{i-1})_++S_i$,  where $T_{i-1}=(S_{i-1}-\tau)_+$. \hfill $\square$

\subsection{Proof of Theorem~\ref{thm:tau}}
\label{p:tau}

We show that for
\[
\tau =\text{ess}\sup(S)-1,
\]
the unique\footnote{The output distributions corresponding to the maximizing input distributions are capacity-achieving output distributions of the corresponding channels and hence are unique (see, for instance,~\cite[Corollary $5.5$]{polyanskiy2024information}).} output distributions corresponding to the maximizing input distributions of the convex optimization problems on either side of~\eqref{eq:strictineq2} are different.

Let $p_i$ denote $p_S(i),~i \ge 0$ and let $q \defeq p_S(\tau+1) > 0$. Then, random variable $$T \defeq (S_1-\tau)_+$$ satisfies
\begin{align}
\mathbb{P}[T = 1] = 1-\mathbb{P}[T = 0] = q. \nonumber
\end{align}
We have
\begin{align}
p_{(X-T)_+}(n) &\defeq \mathbb{P}[(X-T)_+ = n]  \nonumber\\
&= (1-q)\cdot p_X(n) + q \cdot p_X(n+1). \nonumber
\end{align}
For $n \ge 0$, we have
\begin{align}
    w_n &\defeq \mathbb{P}[(X-T)_+ + S_2 = n] \nonumber\\
    &= \sum_{j=0}^n p_j\cdot p_{(X-T)_+}(n-j)  \nonumber\\
    &= \sum_{j=0}^n p_j\cdot\left\{(1-q)\cdot p_X(n-j) + q\cdot p_X(n+1-j)\right\}. \nonumber
\end{align}
Further,
\begin{align}
\mathbb{E}[(X-T)_+] &= \sum_{j=1}^\infty j \cdot p_{(X-T)_+}(j) \nonumber\\
&= \sum_{j=1}^\infty j \cdot \left\{(1-q)\cdot p_X(j) + q\cdot p_X(j+1)\right\}. \nonumber
\end{align}
The Lagrangian with respect to the left-side optimization of \eqref{eq:strictineq2} is given by
\begin{align}
L_1(p_X,\gamma,\delta) = &-\sum_{k=0}^\infty w_k \ln w_k - \gamma\left(\sum_{k=0}^\infty p_X(k) - 1\right)\nonumber\\
&-\delta\left(\mathbb{E}[(X-T)_+] - \left(\frac{1}{\lambda}-\frac{1}{\mu}\right)\right). \nonumber
\end{align}
Taking the derivative with respect to $p_X(n)$ yields
\begin{align}
    &\frac{d}{dp_X(n)}L_1(p_X,\gamma,\delta) = \nonumber\\
    &-\sum_{\ell=-\infty}^\infty \left[(1-q) p_\ell + q  p_{\ell+1}\right]\left(\ln w_{n+\ell} + 1\right)\nonumber\\
    &-\gamma -\delta\left[n(1-q) + (n-1)q\right]. \nonumber
\end{align}
Setting the derivative to zero gives
\begin{align}
&(1-q)\mathbb{E}_S[-\ln w_{n+S}] +q\mathbb{E}_S[-\ln w_{n-1+S}] \nonumber\\
&= 1 + \gamma + \delta (n-q).
\label{eq:wn}
\end{align}
Using the shorthand notation
\begin{align}
x_n &= \mathbb{E}_S[-\ln w_{n+S}],~n \ge 0, \nonumber
\end{align}
yields the recursive equation
\begin{align}
    (1-q)x_n + qx_{n-1} = 1+\gamma + \delta(n-q).
\label{receq}
\end{align}

A particular solution is given by
\begin{align}
x_n^{(\mathrm{p})} = 1+\gamma + n\delta, \nonumber
\end{align}
and the homogeneous solution is given by
\begin{align}
x_n^{(\mathrm{h})} &= \left(\frac{q}{q-1}\right)^n x_0 = \left(\frac{q}{q-1}\right)^n \mathbb{E}_S[-\ln w_S]. \nonumber
\end{align}
Since $x_n = x_n^{(\mathrm{h})} + x_n^{(\mathrm{p})}$, we get
\begin{align}
&\mathbb{E}_S[-\ln w_{n+S}] = 1+\gamma + n\delta  + \left(\frac{q}{q-1}\right)^n\mathbb{E}_S[-\ln w_S]. \nonumber
\end{align}
Plugging in $n=1$ in~\eqref{receq}, we have $$\mathbb{E}_S[-\ln w_S] = 1 +\gamma,$$ which implies
\begin{align}
&\mathbb{E}_S[-\ln w_{n+S}] = 1+\gamma + n\delta  + \left(\frac{q}{q-1}\right)^n(1+\gamma).  \nonumber
\end{align}
Next, we consider the right-side of \eqref{eq:strictineq2}. Given a nonnegative random variable $W$, for $n \ge 0$, we have
\begin{align}
v_n &\defeq \mathbb{P}[W+S=n] \nonumber\\
&= \sum_{j=0}^n p_{W}(n-j) p_j. \nonumber
\end{align}
The Lagrangian is
\begin{align}
L_2(p_W,\alpha,\beta) = &-\sum_{k=0}^\infty v_k \ln v_k - \alpha\left(\sum_{k=0}^\infty p_W(k) - 1\right)\nonumber\\
&-\beta\left(\mathbb{E}[W] - \left(\frac{1}{\lambda}-\frac{1}{\mu}\right)\right). \nonumber
\end{align}
Taking the derivative of $L_2(p_W,\alpha,\beta)$ with respect to $p_W(n)$ yields
\begin{align}
    \frac{d}{dp_W(n)}L_2(p_W,\alpha,\beta) = &-\sum_{\ell=0}^\infty p_\ell \left(\ln v_{n+\ell} + 1\right)-\alpha -n\beta. \nonumber
\end{align}
Setting the derivative to zero gives
\begin{equation}
\mathbb{E}_S[-\ln v_{n+S}] =1 + \alpha + n\beta.
\label{eq:vn}
\end{equation}
Assuming that $w_n = v_n$ yields $\alpha = \gamma$ and comparing \eqref{eq:wn} and~\eqref{eq:vn} for $n=1$ yields $\delta = \beta$, and
therefore
\begin{align}
\mathbb{E}_S[-\ln v_{n+S}] - x_n^{(\mathrm{p})} = 0. \nonumber
\end{align}
However,
\begin{align}
\mathbb{E}_S[-\ln w_{n+S}] - x_n^{(\mathrm{p})} = \left(\frac{q}{q-1}\right)^n(1+\gamma), \nonumber
\end{align}
which is positive for $n$ even and negative for $n$ odd since
\begin{align}
1+\gamma = \mathbb{E}_S[-\ln w_S] > 0. \nonumber
\end{align}
Therefore, we have $w_n \neq v_n$ which completes the proof since the (feedback) capacity-achieving output distribution is unique. \hfill $\square$

\subsection{Proof of Theorem~\ref{thm:ctsbdd}}
\label{s:ctsbdd}
We show that~\eqref{eq:strictineq} is satisfied and use Corollary~\ref{corupt} to conclude. The pdf of $W+S$ is given by
\begin{align}
f_{W+S}(t) = \int_{\mathbb{R}} f_W(t-s)f_S(s) ds. \nonumber
\end{align}
The Lagrangian corresponding to the right-side of~\eqref{eq:strictineq} is given by
\begin{align}
&L_2(P_W,\alpha,\beta)  \nonumber\\
&= -\int_{\mathbb{R}} f_{W+S}(t) \ln f_{W+S}(t) dt - \alpha \left[\int_{\mathbb{R}} dP_W - 1\right] \nonumber\\
&~~~- \beta \left[\int_{\mathbb{R}} t\cdot dP_W - \left(\frac{1}{\lambda} - \frac{1}{\mu}\right)\right]. \nonumber
\end{align}
Let $E$ be the vector space of all signed measures from $\mathbb{R}^+$ to $\mathbb{R}$. By Lebesgue's decomposition theorem (see, for instance,~\cite[Theorem $19.61$]{hewitt1965real}), any $\nu \in E$ can be decomposed uniquely as 
\begin{align}
\nu = \nu_{ac} + \nu_{c} + \nu_{pp}, \nonumber
\end{align}
where $\nu_{ac} \ll Leb_{\mathbb{R}}$ ($Leb_{\mathbb{R}}$ denotes the Lebesgue measure on $\mathbb{R}$), $\nu_c$ is singularly continuous,\footnote{The cdf corresponding to $\nu_{c}$ has derivative (with respect to $\mathbb{R}$) uniformly equal to $0$.} and $\nu_{pp}$ is a discrete measure with pure point masses on a countable set $I$. Define\footnote{Loosely speaking (since $\nu$ is a signed measure), $A_\nu$ denotes the pdf of a random variable $Z+S$ where $Z$ has probability law $\nu$.}
\begin{align}
A_\nu (t) \defeq \int_{\mathbb{R}} \widetilde{\nu}(t-s)f_S(s)ds + \sum_{i \in I} \nu_{pp} (t-i)f_S(i), \nonumber
\end{align}
where $\widetilde{\nu}$ denotes the Radon-Nikodym derivative of $\nu$ with respect to $Leb_{\mathbb{R}}$. With this notation, the Gateaux derivative of $L_2$ at $P_W$ in the direction of $\nu \in E$ is given by
\begin{align}
\delta L_2(P_W,\alpha,\beta; \nu) = &-\int_{\mathbb{R}} A_\nu (t) \Big\{1+\ln A_{P_W}(t)\Big\} dt \nonumber\\
&- \alpha \int_{\mathbb{R}} d\nu - \beta \int_{\mathbb{R}} t d\nu. \nonumber
\end{align}
For $a \in supp(S)$, let $\nu$ be $\delta_{a}$ the Dirac measure centered on $a$. Then,
\begin{align}
A_\nu (t) = A_{\delta_{a}} (t) = f_S(t-a). \nonumber
\end{align}
Thus,
\begin{align}
&\delta L_2(P_W,\alpha,\beta; \nu) \nonumber\\
&= -\int_{\mathbb{R}} f_S(t-a)\Big\{1+\ln A_{P_W}(t)\Big\} dt - \alpha - a\beta  \nonumber\\
&= 0, \nonumber
\end{align}
whereby
\begin{align}
\mathbb{E}_S[-\ln A_{P_W}(S+a)] = 1 + \alpha + a\beta. \nonumber
\end{align}
Setting $a = 0$ yields
\begin{align}
\mathbb{E}_S[-\ln A_{P_W}(S)] = 1 + \alpha. \nonumber
\end{align}
Next we compute the Lagrangian with respect to the left-side of~\eqref{eq:strictineq}. The pdf of $(X-S)_+$ is given by
\begin{align}
f_{(X-S)_+}(t) = b \delta_0(t) + \int_{\mathbb{R}} f_S(s) f_X(t+s) ds \qquad t \ge 0, \nonumber
\end{align}
where
\begin{align}
    b &= \int_{\mathbb{R}} \left(\int_0^a dP_X \right) dP_S. \nonumber
\end{align}
The pdf of $(X-S_1)_+ + S_2$ is given by
\begin{align}
w(t) = b f_S(t) + \int_{\mathbb{R}} \left(\int_{\mathbb{R}} f_S(s)f_X(y+s)ds\right) f_S(t-y) dy  \nonumber
\end{align}
for $t \ge 0$. The Lagrangian with respect to the left-side of~\eqref{eq:strictineq} is given by
\begin{align}
L_1(P_W,\gamma,\epsilon) = &-\int_{\mathbb{R}} w(t) \ln w(t) dt - \gamma \left[\int_{\mathbb{R}} dP_X -1 \right] \nonumber\\
&- \epsilon \left[\int_{\mathbb{R}} t \cdot dP_{(X-S_1)_+}-\left(\frac{1}{\lambda}-\frac{1}{\mu}\right)\right]. \nonumber
\end{align}
The Gateaux derivative of $L_1$ at $P_X$ in the direction of $\nu \in E$ is given by
\begin{align}
\delta L_1(P_X,\gamma,\epsilon; \nu) = &-\int_{\mathbb{R}} B_\nu (t) \Big\{1+\ln B_{P_X}(t)\Big\} dt \nonumber\\
&- \gamma \int_{\mathbb{R}} d\nu - \epsilon \int_{\mathbb{R}} t C_\nu(t) dt, \nonumber
\end{align}
where
\begin{align}
C_\nu (t) \defeq \int_{\mathbb{R}} f_S(s) \widetilde{\nu}(t+s) ds + \sum_{i \in I} f_S(i) \nu_{pp} (t+i) \nonumber
\end{align}
and where
\begin{align}
B_\nu (t) &\defeq \left[\int_{\mathbb{R}} \left(\int_{-\infty}^s d\nu \right) dP_S\right] f_S(t) \nonumber\\
&~~+ \int_{\mathbb{R}} \left(\int_{\mathbb{R}} f_S(s)\widetilde{\nu}(u+s)ds\right) f_S(t-u) du \nonumber\\
&~~+ \int_{\mathbb{R}} \left(\sum_{i \in I} f_S(i)\nu_{pp}(u+i)\right) f_S(t-u) du. \nonumber 
\end{align}
With $\nu = \delta_{a}$ we have
\begin{align}
C_\nu (t) = C_{\delta_{a}}(t) = f_S(a-t) ,\nonumber
\end{align}
and
\begin{align}
B_{\nu}(t) &= B_{\delta_{a}}(t)  \nonumber\\
&= \left\{1-F_S(a)\right\} f_S(t) + \int_{\mathbb{R}} f_S(a-u)f_S(t-u)du. \nonumber
\end{align}
Setting $\delta L_1$ equal to zero yields
\begin{align}
&\left\{1-F_S(a)\right\} \mathbb{E}_S[-\ln B_{P_X}(S)]\nonumber\\
&-\int_{\mathbb{R}} \left(\int_0^{a} f_S(a-u)f_S(t-u)du\right)\ln B_{P_X}(t) dt \nonumber\\
&= 1 + \gamma + \epsilon\int_0^a tf_S(a-t)dt. 
\label{eq:delL1}
\end{align}
Setting $a = 0$ yields
\begin{align}
\mathbb{E}_S[-\ln B_{P_X}(S)] = 1 + \gamma. \nonumber
\end{align}
The double integral in~\eqref{eq:delL1} can be written (by interchanging the order of integration using Fubini's theorem) as
\begin{align}
&-\int_{\mathbb{R}} \left(\int_0^{a} f_S(a-u)f_S(t-u)du\right)\ln B_{P_X}(t) dt \nonumber\\
&= \int_0^{a} f_S(a-u) \left(-\int_{\mathbb{R}} f_S(t-u)\ln B_{P_X}(t)dt\right) du  \nonumber\\
&= \int_0^{a} f_S(a-u) \mathbb{E}_S[-\ln B_{P_X}(S+u)]du. \nonumber
\end{align}
By the same argument, we have
\begin{align}
\int_{\mathbb{R}} \left(\int_0^{a} f_S(a-u)f_S(t-u)du\right) dt = F_S(a). \nonumber
\end{align}
Thus, the two output distributions satisfy, respectively,
\begin{equation}
\mathbb{E}_S[-\ln A_{P_W}(S+a)] = 1 + \alpha + a\beta,
\label{eq:ya}
\end{equation}
and
\begin{align}
&\left\{1-F_S(a)\right\} \mathbb{E}_S[-\ln B_{P_X}(S)] \nonumber\\
&+ \int_0^{a} f_S(a-u) \mathbb{E}_S[-\ln B_{P_X}(S+u)]du \nonumber\\
&= 1  +\gamma + \epsilon\int_0^a tf_S(a-t)dt.
\label{eq:yt}
\end{align}
Suppose the two output distributions match. Then, we have $\alpha = \gamma$. Let $y_t$ denote $\mathbb{E}_S[-\ln A_{P_W}(S+t)]$ for $t \in [0,a]$. Then, identities~\eqref{eq:ya} and~\eqref{eq:yt} can be written as
\begin{align}
y_{a} = 1 + \alpha + a\beta, \nonumber
\end{align}
and
\begin{align}
&\left\{1-F_S(a)\right\} y_{0} + \int_{0}^a f_S(a-t) y_{t}dt  \nonumber\\
&= 1 + \alpha + \epsilon\int_0^a tf_S(a-t)dt. \nonumber
\end{align}
Plugging the first equation into the second yields $\beta = \epsilon$. Thus, the particular solution to the set of equations is given by
\begin{align}
y_t^{(\mathrm{p})} = 1 + \alpha + \beta t. \nonumber
\end{align}
The homogeneous solutions are given by $y(t)$ that satisfy
\begin{align}
\int_{0}^a f_S(a-t)y(t)dt = 0 \nonumber
\end{align}
for all $a \in supp(S)$. Consider the following bounded linear operator on the Banach space $C[0,\infty)$:
\begin{align}
(Ly)(a) = \int_{0}^a f_S(a-t)y(t)dt. \nonumber
\end{align}
If the equation $(Ly)(a) = 0$ has nontrivial homogeneous solutions, then the optimal output distribution under full feedback and weak feedback are different and hence feedback increases capacity. A sufficient condition for the equation to have nontrivial homogeneous solutions is for $L$ to be not invertible. The operator $L$ is a Hilbert-Schmidt operator if
\begin{align}
\int_{0}^{\infty}  \int_{0}^{\infty}  f_S(a-t)^2 dt \cdot da < \infty.
\label{eq:HScondition}
\end{align}
A Hilbert-Schmidt operator is a compact operator and it is well-known that (see, for instance,~\cite[Chapter $6$]{conway1990graduate}) when the domain of a compact operator is infinite-dimensional, then zero belongs to its spectrum. This implies that either $L$ is not invertible or it does not have a bounded inverse. However, since $L$ is bounded, by the inverse mapping theorem~\cite[Theorem $12.5$]{conway1990graduate}, it follows that $L$ is not invertible. Indeed,
\begin{enumerate}
    \item[i.] When $supp(S) = [a,b]$ where $a > 0$ and $b < \infty$,~\eqref{eq:HScondition} holds and hence $L$ is Hilbert-Schmidt.

    \item[ii.] When $supp(S) = [a,b]$ where $a \ge 0$ and $b < \infty$ and $f_S(\cdot)$ is continuous on $[a,b]$,~\eqref{eq:HScondition} holds and hence $L$ is Hilbert-Schmidt.
\end{enumerate}

\begin{rem}
     When $S$ has exponential distribution with mean $1/\mu$, we have
    \begin{align}
    \int_{0}^{\infty} \int_{0}^{\infty} f_S(a-t)^2 dt \cdot da &= \int_{0}^{\infty} \int_{0}^{\infty} \mu e^{-2\mu(a-t)} da \cdot dt  \nonumber\\
    &= \infty \nonumber
    \end{align}
    and hence $L$ is not Hilbert-Schmidt. Indeed, for this case, $C_{\mathrm{WF}} = C_{\mathrm{F}}$.\footnote{Equality is achieved by $X$ distributed as
\begin{align}
   \mathbb{P}[X=0] &= \beta, \nonumber\\
   f_X(x) &= (1-\beta)\lambda e^{-\lambda x}; ~~x > 0, \nonumber
\end{align}
where $\beta = {\lambda^2}/{\mu^2}$ and $W$ as exhibited in~\cite[Theorem $3$]{anantharam1996bits}.}
\hfill $\square$
\end{rem}

\subsection{Proof of Theorem~\ref{thm:inf} -- discrete-time setting}
\label{p:discreteunbdd}
We show that the optimization problems on either side of~\eqref{eq:eqinf} have the same solution. This is established by showing that the non-homogeneous recurrence relation with variable coefficients resulting from the KKT conditions has no homogeneous solution and that the unique particular solutions match.

For 
\begin{align}
v_n \defeq \mathbb{P}[W+S = n], n \ge 0, \nonumber
\end{align} 
we have (see~\eqref{eq:vn} in the proof of Theorem~\ref{thm:tau}),
\begin{equation}
\mathbb{E}_S[-\ln v_{n+S}] =1 + \alpha + n\beta.
\label{eq:weakvn}
\end{equation}
Fix $0 \le \tau < \infty$. Let $p_i$ denote $P_S(i), i \ge 0$ and
\begin{align}
    q_n &\defeq \mathbb{P}[T = n] \nonumber\\
    &= \begin{cases}
    \sum_{k=0}^\tau p_k &\text{if } n=0 \nonumber\\
    p_{n+\tau} &\text{if } n \ge 1. \nonumber
    \end{cases}
\end{align}
For simplicity, assume that $q_0 > 0$.\footnote{If $q_0 = 0$, all the arguments follow almost verbatim with $q_j, q_{j+1},\ldots$ in place of $q_0, q_1, \ldots$ where 
\begin{align}
j = \min\{i > 0:q_i > 0\}. \nonumber
\end{align}}   
Then,
\begin{align}
    p_{(X-T)_+}(n) &\defeq \mathbb{P}[(X-T)_+ = n]  \nonumber\\
    &= \begin{cases}
    \sum_{k=0}^\infty \left(q_k \sum_{u=0}^k p_X(u)\right) &\text{if } n=0 \nonumber\\
    \sum_{i=0}^\infty q_i p_X(n+i) &\text{if } n \ge 1.
    \end{cases}
\end{align}
Therefore,
\begin{align}
\mathbb{E}[(X-T)_+] &= \sum_{j=1}^\infty j \left(\sum_{i=0}^\infty q_i p_X(j+i)\right), \nonumber
\end{align}
and
\begin{align}
    &w_n \defeq \mathbb{P}[(X-T)_+ + S_2 = n] \nonumber\\
    &~~~~=\sum_{j=0}^n p_{(X-T)_+}(n-j) p_j \nonumber\\
    &= \sum_{k=0}^\infty \left(q_k \sum_{u=0}^k p_X(u)\right)p_n + \sum_{j=1}^{n} \left(\sum_{i=0}^\infty q_i p_X(j+i)\right)p_{n-j}. \nonumber
\end{align}
The Lagrangian with respect to the left-side of~\eqref{eq:eqinf} is given by
\begin{align}
L_1(p_X,\gamma,\delta) = &-\sum_{k=0}^\infty w_k \ln w_k - \gamma\left(\sum_{k=0}^\infty p_X(k) - 1\right)\nonumber\\
&-\delta\left(\mathbb{E}[(X-T)_+] - \left(\frac{1}{\lambda}-\frac{1}{\mu}\right)\right). \nonumber
\end{align}
The derivative of $L_1(p_X,\gamma,\delta)$ with respect to $p_X(n)$ is 
\begin{align}
    &\frac{d}{dp_X(n)}L_1(p_X,\gamma,\delta) = -\sum_{\ell=-\infty}^\infty \Big\{(q_n+q_{n+1}+\ldots)p_{n+\ell} \nonumber\\
    &+ q_{n-1}p_{n+\ell-1} + \ldots + q_1p_{\ell+1} + q_0p_\ell\Big\}\left(\ln w_{n+\ell} + 1\right)\nonumber\\
    &-\gamma -\delta\Big(\sum_{j=1}^njq_{n-j}\Big). \nonumber
\end{align}
Let 
\begin{align}
x_{n} \defeq \mathbb{E}_S[-\ln w_{n+S}]. \nonumber
\end{align}
Setting the derivative of $L_1(p_X,\gamma,\delta)$ with respect to $p_X(n)$ to zero yields, for $n \ge 1$,
\begin{align}
&q_0x_n + q_1x_{n-1} + \ldots +  q_{n-1}x_{1} + \Big(1- \sum_{i=0}^{n-1}q_i\Big)x_0 \nonumber\\
&= 1 + \gamma + \delta\Big(\sum_{j=1}^njq_{n-j}\Big)
\label{eq:recursion}
\end{align}
and $x_0 = 1+\gamma$. This is an order $n$ non-homogeneous recurrence relation with variable coefficients~\cite[Chapter $2$]{greene1990mathematics}. It can be checked (by comparing it with~\eqref{eq:weakvn} for $n=0$ and $n=1$) that $\alpha = \gamma$ and $\beta = \delta$. Further, it can be checked (by substitution) that the particular solution to the recurrence relation in~\eqref{eq:recursion} is given by 
\begin{align}
x_n^{(\mathrm{p})} = 1+\gamma+n\delta \nonumber
\end{align}
which matches \eqref{eq:weakvn}. The homogeneous solutions to the recurrence relation in~\eqref{eq:recursion} are obtained by solving the corresponding homogeneous recurrence relation, namely for $n \ge 1$,
\begin{equation}
q_0x_n + q_1x_{n-1} + \ldots +  q_{n-1}x_{1} + \Big(1- \sum_{i=0}^{n-1}q_i\Big)x_0 = 0.
\label{eq:homo}
\end{equation}
Since $x_0 = 1+\gamma$, the recurrence relation in~\eqref{eq:recursion} can be written as
\begin{align}
Ax = b \nonumber
\end{align}
where $A$ is an infinite Toeplitz matrix~\cite{bottcher2012toeplitz} that is the limit of the sequence of $n \times n$ Toeplitz matrices $A_n$ given by
\begin{align}
A_n = \begin{bmatrix}
    q_0 & 0 & 0 & \cdots & 0\\
    q_1 & q_0 & 0 & \ddots & \vdots\\
    q_2 & q_1 & q_0 & 0 & \ddots\\
    \vdots & \ddots & \ddots & \ddots & \ddots\\
    q_{n-1} & q_{n-2} & \cdots & q_1 & q_0
\end{bmatrix}, \nonumber
\end{align}
and 
\begin{align}
x &= [x_1-(1+\gamma)~~x_2-(1+\gamma)~~x_3-(1+\gamma)~~\cdots~]^T, \nonumber\\
b &= [\delta q_0~~\delta (q_1+2q_0)~~\delta (q_2+2q_1+3q_0)~~\cdots~]^T \nonumber
\end{align}
with $[\cdots]^T$ denoting the transpose. Finding solutions to the homogeneous recurrence relation in~\eqref{eq:homo} is equivalent to finding solutions $x^*$ such that $Ax^* = 0$. By~\cite[Section $6.4$, $4.2$]{bottcher2012toeplitz}, the system $Ax = 0$ has a non-zero solution if and only if 
\begin{align}
\lim_{n \to \infty} \inf \sigma_{\min}(A_n) = 0, \nonumber
\end{align}
where $\sigma_{\min}(A_n)$ denotes the smallest singular value of $A_n$. It can be seen that the $A_n$'s are lower triangular with all diagonal entries equal to $q_0$ and hence for all $n \ge 1$, we have
\begin{align}
\sigma_{\min}(A_n) = q_0>0. \nonumber
\end{align} 
Thus, there are no homogeneous solutions to the recurrence relation in~\eqref{eq:recursion}. Hence, the solutions to the optimization problems on either side of~\eqref{eq:eqinf} are the same. This completes the proof.\hfill $\square$

\subsection{Proof of Theorem~\ref{thm:inf} - continuous-time setting}
\label{p:ctsunbdd}

We assume that~\eqref{eq:eqinf} does not hold for an unbounded continuous-time service time distribution $S$ and arrive at a contradiction to Theorem~\ref{thm:inf} under the discrete-time setting, proved in Section~\ref{p:discreteunbdd}.

Throughout this proof, we use $h(\cdot)$ to denote differential entropy and $H(\cdot)$ to denote discrete entropy. Convergence of random variables is in distribution. Suppose $X$ and $W$ achieve the left- and right-sides of~\eqref{eq:eqinf}, respectively. Now, suppose that
\begin{equation}
h(W+S) = h((X-T)_+ + S_2) + \epsilon
\label{eq:heps}
\end{equation}
for some $\epsilon > 0$. We will show that this results in a contradiction. For a random variable $Z$, let $Z^\Delta$ denote the discrete random variable corresponding to the uniform quantization of $Z$ with interval width $\Delta$. Clearly, $Z^\Delta \to Z$ as $\Delta \to 0$. The differential entropy of $Z$ is defined as~\cite[Section $7.4$]{gray2011entropy}
\begin{align}
h(Z) = \sup_{\Delta} H(Z^\Delta). \nonumber
\end{align}

For $\Delta > 0$, let $\widetilde{X}^\Delta$ be the input distribution that achieves\footnote{For simplicity, throughout we assume that the distributions achieve the supremum and hence the maximum. This does not change the results---indeed, we can choose $\delta$ and $\Delta$ appropriately to accommodate the difference.} the upper bound on the weak feedback capacity for service time distribution $S^\Delta$. Similarly,  let $\widetilde{W}^\Delta$ be the input distribution that achieves feedback capacity for service time distribution $S^\Delta$. Let 
\begin{align}
Y \defeq (\widetilde{X}-T)_+ + S_2. \nonumber
\end{align}
Since, $Y^\Delta \to Y$ and also
\begin{align}
(\widetilde{X}^\Delta - T^\Delta)_+ + S_2^\Delta \to Y, \nonumber
\end{align}
for $\delta < \epsilon$, we can choose $\Delta$ small enough so that
\begin{align}
 H((\widetilde{X}^\Delta - T^\Delta)_+ + S_2^\Delta) \le H(Y^\Delta) + \frac{\delta}{2}. \nonumber
\end{align}
Then,
\begin{align}
 H((\widetilde{X}^\Delta - T^\Delta)_+ + S_2^\Delta) + \frac{\delta}{2}
 &\le H(Y^\Delta) + \delta \nonumber\\
 &\le h((\widetilde{X}-T)_+ + S_2) + \delta \nonumber\\
 &\le h((X-T)_+ + S_2) + \delta \nonumber\\
 &< h(W+S), \nonumber
\end{align}
where the last inequality follows from~\eqref{eq:heps} and $\delta < \epsilon$. Again, since $(W+S)^\Delta \to W+S$ and $W^\Delta + S^\Delta \to W+S$, we have
\begin{align}
h(W+S) \le H(W^\Delta + S^\Delta) + \frac{\delta}{2} \nonumber
\end{align}
for $\Delta$ small enough. Therefore,
 \begin{align}
 H((\widetilde{X}^\Delta - T^\Delta)_+ + S_2^\Delta) + \frac{\delta}{2}
  &< H(W^\Delta + S^\Delta) + \frac{\delta}{2} \nonumber\\
 &\le H(\widetilde{W} + S^\Delta)  + \frac{\delta}{2}, \nonumber
\end{align}
whereby we have
\begin{align}
H((\widetilde{X}^\Delta - T^\Delta)_+ + S_2^\Delta) < H(\widetilde{W} + S^\Delta). \nonumber
\end{align}
This is a contradiction to the discrete counterpart of the theorem (proved in Section~\ref{p:discreteunbdd}) which states that for $S^\Delta$ a discrete-valued service time with infinite support, the two bounds match. This concludes the proof.\hfill $\square$

\section{Concluding remarks}
\label{s:conclusion}
We have established that full feedback increases the capacity of FIFO queues with bounded service times. This result was obtained by investigating generalized feedback, which interpolates between weak and full feedback.

For the case of unbounded service times the problem of characterizing the service times for which full feedback increases capacity remains open. We have no examples of specific service times with unbounded support for which feedback increases capacity, and our general approach is inconclusive as all the derived bounds are equal to the full-feedback capacity (Theorem~\ref{thm:inf}). One possible reason for this is that our upper bound on the weak feedback capacity (Proposition~\ref{prop:CWF_UB}) is not tight. As noted in the remark following Proposition~\ref{prop:CWF_UB}, the outputs $\{W_i(X_i)+S_i\}_{i=1}^n$ are generally dependent since $W_i(X_i)=(X_i-S_{i-1})_+$ depends on $S_{i-1}$ and it may be possible to derive a tighter upper bound via a more careful analysis of the output entropy $H(\{W_i(X_i)+S_i\}_{i=1}^n)$.

\section*{Acknowledgements}
KRS thanks Fran\c{c}ois Baccelli and Farzan Farnia for useful discussions.



\end{document}